\documentclass[reprint,superscriptaddress,amsmath,amssymb,aps,prb,showpacs]{revtex4-1}
\usepackage{graphicx}
\usepackage{dcolumn}
\usepackage{bm}
\usepackage{hyperref}
\usepackage{multirow}
\usepackage{tabularx}
\usepackage{siunitx}

\usepackage[utf8]{inputenc}

\begin{document}

\title{Applicability of the Debye-Waller damping factor for the determination of the line-edge roughness of lamellar gratings}

\author{Anal\'{i}a Fern\'{a}ndez Herrero}
\affiliation{Physikalisch-Technische Bundesanstalt (PTB), Abbestr. 2-12, 10587 Berlin, Germany}
\author{Mika Pflüger}
\affiliation{Physikalisch-Technische Bundesanstalt (PTB), Abbestr. 2-12, 10587 Berlin, Germany}
\author{J\"{u}rgen Probst}
\affiliation{Helmholtz-Zentrum Berlin (HZB), Albert-Einstein-Str. 15, 12489 Berlin, Germany}
\author{Frank Scholze}
\affiliation{Physikalisch-Technische Bundesanstalt (PTB), Abbestr. 2-12, 10587 Berlin, Germany}
\author{Victor Soltwisch}
\affiliation{Physikalisch-Technische Bundesanstalt (PTB), Abbestr. 2-12, 10587 Berlin, Germany}

\begin{abstract}
Periodic nanostructures are fundamental elements in optical instrumentation as well as basis structures in integrated electronic circuits. Decreasing sizes and increasing complexity of nanostructures have made roughness a limiting parameter to the performance. Grazing-incidence small-angle X-ray scattering is a characterization method that is sensitive to three-dimensional structures and their imperfections. To quantify line-edge roughness, a Debye-Waller factor (DWF), which is derived for binary gratings, is usually used. In this work, we systematically analyze the effect of roughness on the diffracted intensities. Two different limits to applying the DWF are found depending on whether or not the roughness is normally distributed.
\end{abstract}

\maketitle
\section{Introduction}

Demand for smaller functional nanostructures has driven the development of new metrology tools that can resolve three-dimensional structures while addressing the collateral effects of shrinking the dimensions (i.e., additional imperfections). 
Roughness can be produced at different steps of the manufacturing process, and becomes increasingly relevant as the dimensions of the nanostructures are reduced. In the case of line-shaped nanostructures (e.g., gratings), the most characteristic type of roughness is along the edges. Line-edge roughness refers to the randomly distributed variation of the edges of a line. This kind of roughness is highly relevant when producing lamellar nanostructures. For instance, in the semiconductor industry because it results in a device-to-device mismatch and is one of the most critical random variation sources~\cite{variation_shin_2016}. In most cases, the characterization in the sub-nm range of the feature sizes and shapes has been addressed in several methods, of which atomic force microscopy (AFM) and critical dimension scanning electron microscopy (CD-SEM) are commonly used~\cite{patsis_roughness_2003, variation_shin_2016, li_noise_2016, bunday_hvm_2016}. However, such methods are of limited use when applied to the characterization of nanostructures. As the complexity of the nanostructures increases, it becomes very difficult to access parameters such as the height or the sidewall angle by means of a scanning method~\cite{Roman_height_2004}. For this reason, a total characterization of the line-shape is either not possible or subject to the development of a more complex data analysis~\cite{lane_global_2017,sun_line_2004}. Even if a more robust analysis is performed, a large segment of the device area must be mapped in order to obtain statistically significant results for parameters like roughness. A detailed analysis of the roughness would significantly increase the measurement times.

In contrast, X-ray scattering techniques are a rapid measurement alternative that can provide statistical information in a very short time. A well-known method for the characterization of nanostructured surfaces is scattering under grazing incidence angles close to the total external reflection. An advantage of grazing incidence small-angle X-ray scattering (GISAXS) is that it allows macroscopic areas to be inspected while also yielding nanoscopic information. GISAXS shows high sensitivity to the surface, as well as to the normal and lateral structures and their imperfections. Therefore, the intensity distribution obtained by GISAXS is sensitive to the cross section of the nanostructures as well as to the imperfections of the structures~\cite{mikulik_coplanar_2001, holy_grazing_2001, soltwisch_correlated_2016,Sawa_importance_2016, suh_characterization_2016}. An imprint of the roughness in the diffuse scattering background and a resulting intensity loss in the diffraction orders have also been reported~\cite{torcal-milla_diffraction_2008, kato_effect_2010, bilski_about_2011, kato_analytical_2012,gross_modeling_2012, suh_characterization_2016}. The scattered intensity in the diffraction orders is usually used to reconstruct the line shape of the nanostructures. Therefore, it is of high interest to have a method that can quantify roughness and be introduced into the reconstruction process. \\

Several studies on the determination of the dimensional parameters of lamellar gratings have identified the roughness as a key parameter for the characterization of nanopatterned structures ~\cite{gross_modeling_2012,  henn_improved_2014}. Both experimental and theoretical reports have indicated the usability of a damping factor to account for the roughness that affects the intensity of the higher diffraction orders. The damping factor is known as the Debye-Waller factor and has been widely used in literature~\cite{erko_multilayer_1993,wang_line_2007,wang_characterization_2007,lemaillet_intercomparison_2013,Levi_holistic_2018}. This factor was analytically derived for binary gratings with line roughness~\cite{kato_effect_2010} and investigated concerning theoretical distributions of roughness ~\cite{wang_characterization_2007, gross_investigations_2010, gross_impact_2017}. It was even included in the reconstruction of state-of-the-art lamellar gratings from GISAXS experiments ~\cite{soltwisch_reconstructing_2017, wang_small_2007}. Therefore, it is necessary to investigate the limits to which the DWF can be applied.\\

To study the suitability of the Debye-Waller factor (DWF) for describing the line roughness of periodic nanostructures, we have systematically analyzed the impact of roughness on the GISAXS scattering pattern. A set of gratings, in which the amplitude of the roughness was systematically changed, was designed. The gratings were measured using SEM to deliver the reference values for the edge placement distributions obtained. Likewise, they were measured using GISAXS for the analysis of the applicability of the DWF. For the computation of the diffraction efficiencies without roughness impact, we used a Maxwell solver based on the finite element method. For the reconstruction of the line shape and validation of the parameter uncertainties, we used the Markov Chain Monte Carlo sampling technique~\cite{Foreman_emcee_2013}.

\section{Experimental setup}

The experiments were conducted at PTB using the four-crystal monochromator (FCM) beamline~\cite{krumrey_high-accuracy_2001} at the electron storage ring BESSY II.  This beamline covers a photon energy range from 1.75 keV to 10 keV. A beam-defining pinhole of about 500 $\mu$m was used at a distance of about 1.5 m to the sample. Together with a scatter guard of 1000 $\mu$m close to the sample, the beam spot size was about 0.5 mm $x$ 0.5 mm at the sample position. The beam had a horizontal divergence of 0.01$^\circ$ and a vertical divergence of 0.006 $^\circ$.

\begin{figure}
    \centering
    \includegraphics[width=.7\linewidth]{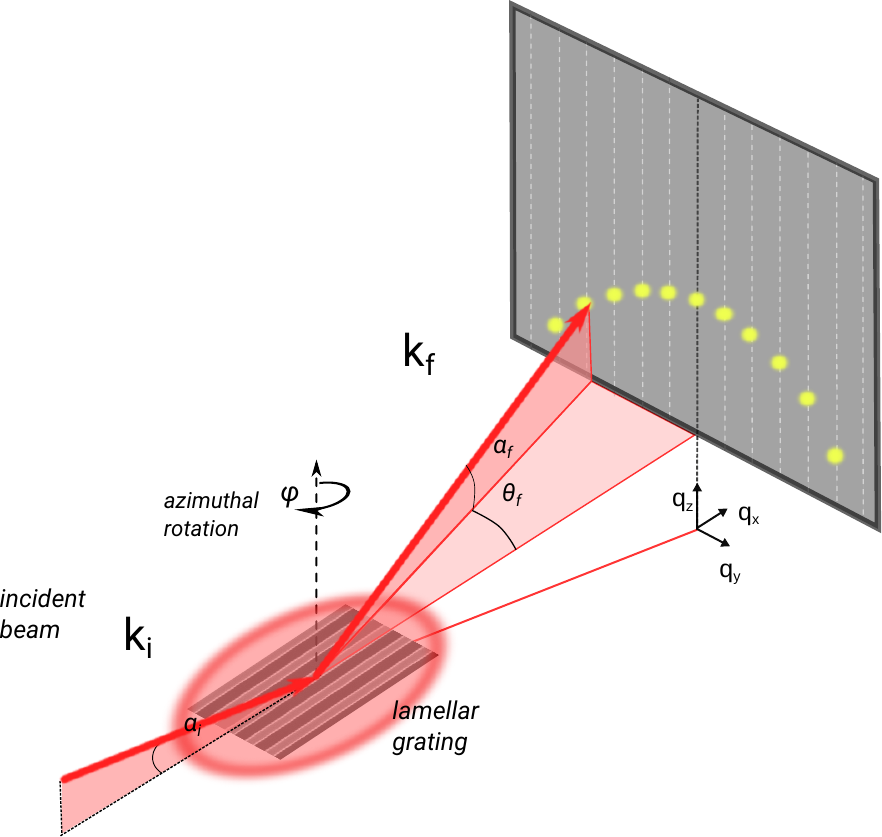}
    \caption{Sketch of the experimental setup. An incoming wave of wavelength $\lambda$ impinges on the sample at an angle $\alpha_i$ and is scattered under the angles $\theta_f$ and $\alpha_f$. The red circle around the grating target represents the illuminated area.}
    \label{fig:figure_sketch}
\end{figure}

The experimental setup is illustrated in Fig.~\ref{fig:figure_sketch}. A monochromatic X-ray beam with a wave vector $\vec{k}_i$ impinges on the sample surface at an incidence angle $\alpha_i$. The elastically scattered wave vector $\vec{k}_f$ propagates with an exit angle $\alpha_f$ and an azimuthal angle $\theta_f$. The sample is placed in a goniometer inside an ultrahigh vacuum chamber that can be rotated around its surface normal at an angle $\varphi$.  

The coordinates in reciprocal space correspond to the momentum transfer,

\begin{equation}
\begin{pmatrix}

q_x \\ 
q_y \\ 
q_z  
\end{pmatrix}
= \frac{2\pi}{\lambda}
\begin{pmatrix}
 \cos(\theta_f) \cos(\alpha_f) - cos(\alpha_i)\\
 \sin(\theta_f) \cos(\alpha_f)\\
\sin(\alpha_f) + \sin(\alpha_i)
\end{pmatrix}.
\label{eq_q}
\end{equation}

\noindent 
The diffraction orders are given by the intersection of the Ewald's sphere with the reciprocal lattice of the grating. If the conical diffraction geometry is chosen, the projection of the incidence plane is parallel to the grating lines, $\varphi=0$. Then, the diffraction orders describe a semicircle in the detector plane~\cite{mikulik_coplanar_2001,yan_intersection_2007} and the position of the orders corresponds to $q_y=m\frac{2\pi}{pitch}$, where m $\in \mathbb{N}$ is the order of diffraction. The azimuth angle $\varphi$ was aligned in such a way that this condition was met, with the elevation angle from the sample horizon of the respective positive and negative diffraction orders being equal. The angular uncertainty achieved in $\varphi$ was $0.01 ^\circ$. 

The detector is an in-vacuum PILATUS 1M detector~\cite{wernecke_characterization_2014} with a pixel size of $(172$ x $172)$ $\mu$m$^2$, placed at a distance of about 3.5 m. The incidence angle is approximately $\alpha_i = 0.8^\circ$. In order to obtain more information on the structures, several parts of the reciprocal space were mapped by varying the photon energy.

\section{Sample set}

A set of samples was produced by means of e-beam lithography at the Helmholtz-Zentrum Berlin. They were etched in a Si wafer. A set composed of five samples was produced: one reference grating where no additional roughness was introduced and four gratings with a dedicated roughness pattern. The area of each of the gratings is 0.5 mm by 4 mm, with the lines extending along the longer direction. The rough gratings were produced to systematically analyzed the effect of the edge roughness in the scattering pattern. 

Two forms of line roughness can be distinguished according to the correlation between the two edges of the line: line edge roughness (LER) and line width roughness (LWR). In the first case, the center position of the line varies along the line while the line width is kept constant (the total correlation between the edges of each line). In the latter case (LWR), the width varies and the position of the line remains constant (the two edges of the line are anti-correlated). 
  
\begin{figure}
     \centering
         \includegraphics[width=.8\linewidth]{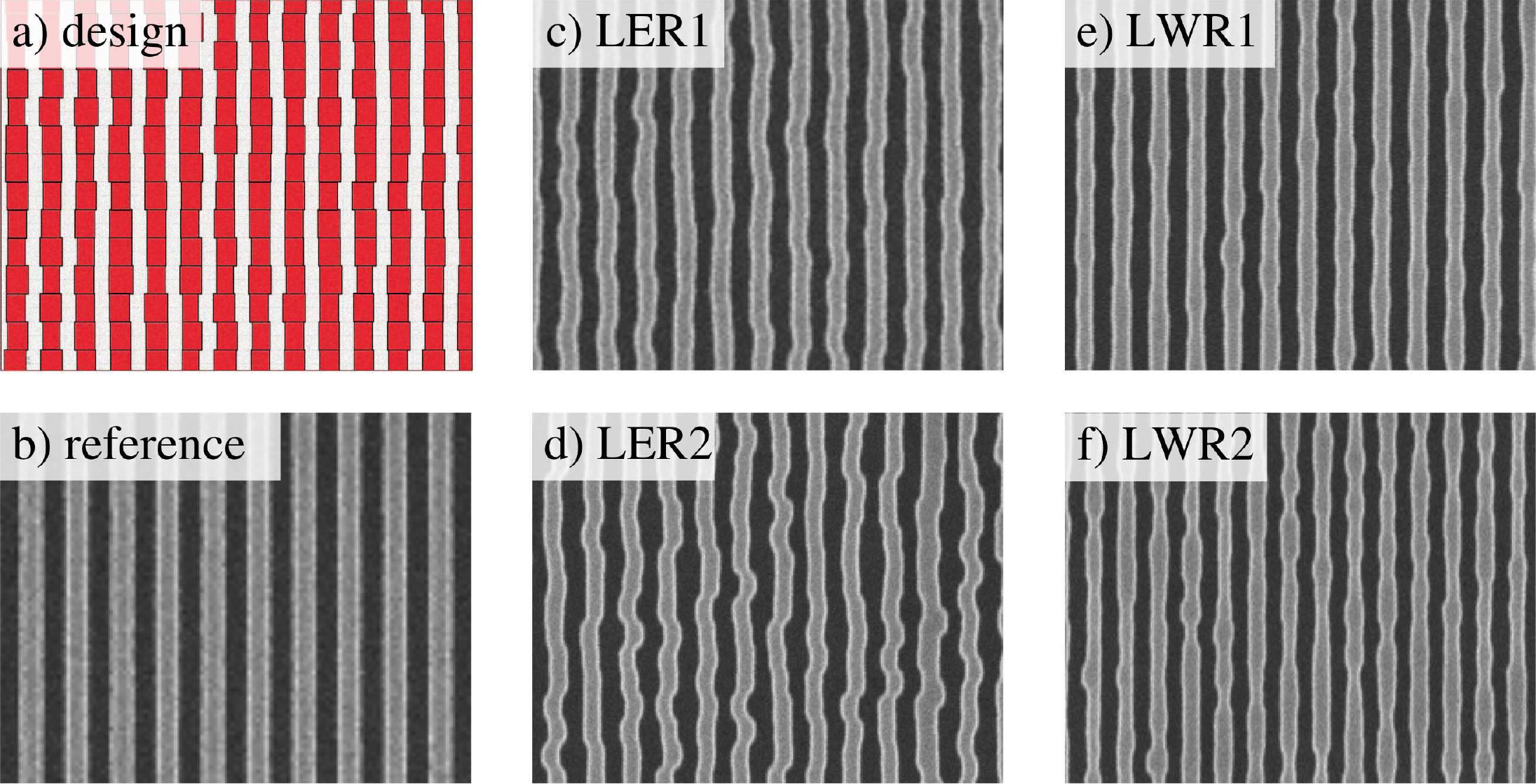}
        \caption{Design and SEM images of the targets. a) shows the design layout of the structure with LER and $\delta_{max}=10 $ nm. The red boxes show the part to be etched. In images b) trough f), the SEM images from one part of the targets are shown: b) is the reference grating, c) LER1: structure with LER and $\delta_{max}$ = 10 nm, d) LER2: LER and $\delta_{max}$ = 20 nm, e) LWR1: LWR and $\delta_{max}$ = 10 nm, and f) LWR2: LWR and $\delta_{max}$ = 20 nm.}
        \label{fig:sem}
\end{figure}

The parameters of the designed structures were chosen in such a way that big amplitudes of the line roughness can be achieved. The samples must have a type of line roughness that predominates over the natural roughness. At the same time, the grating lines must be well-resolved and must not collide. We also tried to avoid other types of roughness (e.g., surface, height) by studying the lithography and etch process windows with line and space width variations.
A pitch of 150 nm and a line width of 65 nm were chosen to satisfy these requirements. Each line can be interpreted as a line made up of boxes of 100 nm by 65 nm, with the longer side along the line, and a height of 120 nm. The size and position of these boxes are changed for the samples with roughness. For the roughness design, a squared basis cell of  51 $\mu$m side-length was designed to limit the data volume for the writing process of the stochastic lines. An example of the design is shown in Figure~\ref{fig:sem} a), where the red colored boxes represent the part to be etched. 
The periodicity of the edge waviness is 100 nm, which is the same for all rough gratings. The position or width of these boxes varies, according to a LER or a LWR design and following a discrete uniform distribution. Each type of roughness (LER/LWR) is covered by two gratings with a different maximum amplitude of the distribution. The two maximum perturbation amplitudes were set to 10 nm and 20 nm. In the case of LWR, the maximum perturbations of the box width translate into a maximum edge displacement $\delta_{max}= 5$ nm or $\delta_{max}= 10$ nm for each sample. The e-beam writer's resolution was set to 1 nm; and therefore, all positions of a given line edge were located within a 1 nm grid. As a result, the discrete steps of the distribution for the LER are 1 nm, whereas for the LWR is 2 nm (as a minimum of 1 nm varies from each box side). Figure~\ref{fig:sem} a) shows the design layout for a sample with line edge roughness and $\delta_{max} = 10$ nm. Figure~\ref{fig:sem} b) shows the SEM image from the reference grating, while in images c) trough f), the SEM images of the rough gratings are shown.

\subsection{Pattern transfer analysis}
The fidelity of the patterns transferred during the etch process are analyzed to compare them with the roughness values obtained with GISAXS.
In our approach, we have catalogued only one type of roughness and addressed it in the form of blocks. Figure~\ref{fig:sem} a) shows the part to be etched in a grid of 1 nm as red boxes. However, the SEM image of the etched gratings (see Figure~\ref{fig:sem} c)) shows a smoother transition between the blocks than the design  because the lithography and etching act as low-pass filters. We have analyzed the SEM images in order to quantify the roughness distribution actually obtained.

\begin{figure}
    \centering
    \includegraphics[width=\linewidth]{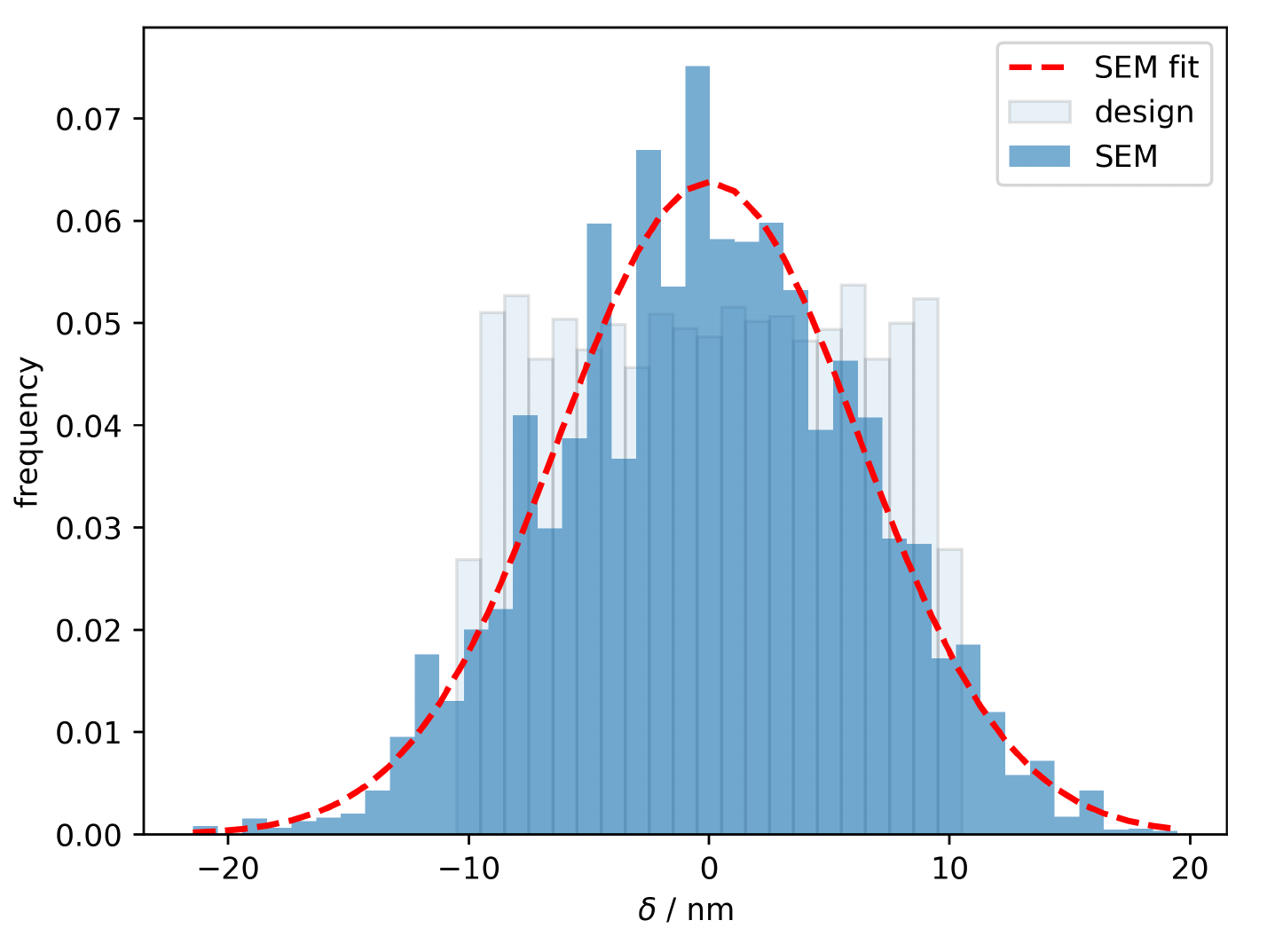}
     \caption{Pattern transfer analysis. Comparison between the design and the etched structures for the grating with line edge roughness and $\delta_{max}$ = 10 nm. The roughness distribution of the design (light blue) is compared to the roughness distribution of etched area (darker blue) obtained in the SEM analysis. The red curve shows the Gauss-curve corresponding to the variance of the distribution obtained.}
     \label{fig:comparison_sem}
\end{figure}

The estimation of the structural parameters defining the line profile from the SEM images usually involves complex data analysis. However, in this case, our aim was merely to analyze the roughness obtained after the etching process of the designed structures. Therefore, a technique for detecting the edges is used and repeated for all SEM images. The edge position of a line in an SEM image is represented by a high variation of the intensity. This intensity variation is used to estimate the boundaries of the line in terms of pixels. 
The displacement of the edges from an ideal straight line is determined. To obtain better statistics, several images from different regions were used.

Figure~\ref{fig:comparison_sem} shows a comparison between the distribution obtained and the target distribution. The discrete uniform distribution design is not conserved after the etching process; instead, a more Gauss-shaped distribution is observed. In other words, the convolution of the roughness distribution introduced and other roughness sources leads to another distribution. We can assume that the distribution of the natural roughness of the samples written by means of e-beam matches that of a Gaussian profile, as it results by the superposition of different and independent sources of roughness (as stated by the central limit theorem). The standard deviation of the distribution, $\xi_{SEM}$, is considered the nominal roughness value. This value is compared to the values obtained from the scattering data.

\section{Impact of imperfections on the scattering pattern}\label{sec:DWF}
\begin{figure*}
  \centering
     \includegraphics[width=\linewidth]{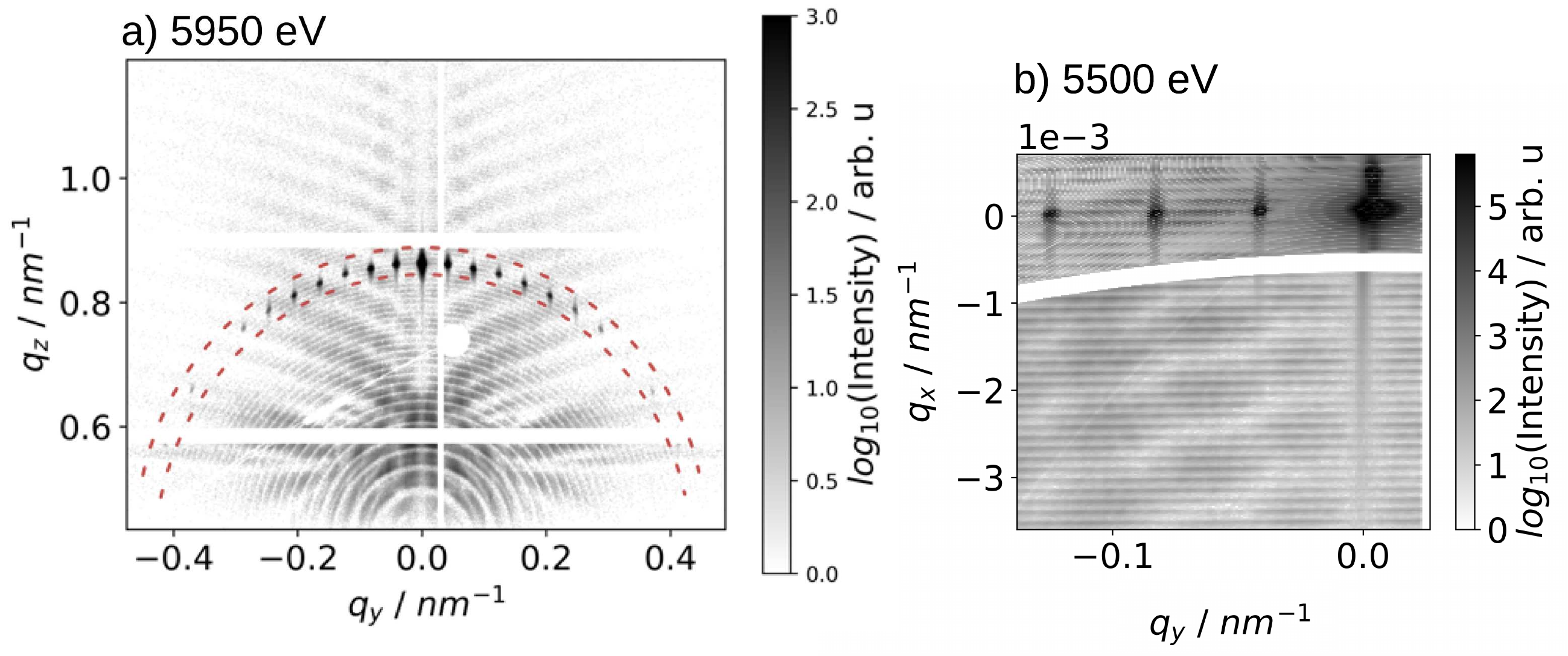}
  \caption{ Scattering pattern from the LER1 grating. a) Scattering pattern with the diffraction orders between the two dashed red lines. Outside of this area, other scattering effects are visible. b) Closer view of one part of the scattering pattern in the $q_x$-$q_y$ scattering plane, where parallel lines in $q_x$ are visible. The distance between these lines corresponds to the 51 $\mu$m side length of the squared basis cell used for the design of the roughness.}
      \label{fig:GISAXS}
\end{figure*}

In the presence of roughness, light is scattered out of the position of the diffraction orders and several scattering phenomena can be observed~\cite{suh_characterization_2016,soltwisch_correlated_2016, Fernandez_characteristic_2017, Heusinger_diffuse_2018}. The total intensity of the out-scattered light depends on the amplitude of the roughness. Although this stray light could deliver further information on the roughness and the nano-structure itself, in most cases, only the regular diffraction orders are considered for the reconstruction.

Roughness damps the intensity of the main diffraction orders and must be considered in the characterization of the structures~\cite{germer_modeling_2007,gross_investigations_2010, Endres_investigations_2014, Henn_improved_2012, suh_characterization_2016}. Several reports have examined the role of different types of roughness by calculating their impact on the scattering pattern~\cite{kato_effect_2010, gross_modeling_2012, wang_small_2007}. In the case of lamellar gratings, for the sake of simplicity, a binary grating with a Gaussian distribution of roughness has been considered. Ultimately, a similar form of a Debye-Waller-like factor (DWF) is obtained~\cite{kato_effect_2010}. However, the applicability of this factor for three-dimensional structures has not been crosschecked. In other words, results obtained using the Debye-Waller factor from a GISAXS pattern was assumed to represent the real roughness of the edge~\cite{soltwisch_reconstructing_2017} without further testing of this theory on true three-dimensional structures.

The DWF is well known from crystallography and accounting for the effect of the thermal motion of the atoms in a crystal. Here, the factor has the same form and accounts for the effect of position variations of the edge along the lines.  For a certain lattice size (in this 1-D case, the $pitch$) the boxes oscillate with a value $\delta$ around the equilibrium position. The displacements are normally distributed in such a way that the mean <$\delta$> = 0. This results in a damping~\cite{kato_effect_2010} of the scattered intensity, following
 
\begin{equation}\label{eq:dwf}
\begin{split}
 I_{DWF}(q_y) & = I_0(q_y)\exp{(-\langle\delta^2 \rangle q_y^2)}\\ 
& = I_0(q_y)\exp{(-\xi^2 q_y^2)},  
\end{split}
\end{equation}

\noindent where $I_0(q_y)$ is the intensity of the diffraction orders of the undisturbed grating at $q_y$ and $\xi$ is the roughness parameter; thus, $\xi^2$ corresponds to the variance of the distribution of $\delta$. A short derivation of the DWF for LER is included in the appendix for a better comprehension of the limits of application of the DWF.

It should be noted that, for the design, two types of roughness were distinguished: line edge roughness and line width roughness. However, the Debye Waller factor accounts only for the effective roughness of the edge and not for the correlation of the edges. To distinguish the roughness types, we previously reported the observation of the resonant diffuse scattered sheets~\cite{Fernandez_characteristic_2017}.

Figure~\ref{fig:GISAXS} shows the light scattered from the sample with a low amplitude of the line edge roughness (LER1). The diffraction orders used for the reconstruction are shown between the red dashed lines. Concentric semicircles appear in the diffuse scattering background. For a better view of the diffuse scattered light, a scattered pattern measured at a different photon energy is shown in the $q_y$-$q_x$ scattering plane in Figure~\ref{fig:GISAXS} b). The parallel lines are repeated at a constant distance $q_x = \frac{2\pi}{d_{basis}}$, where $d_{basis} = 51 \mu$m is the side length of the square basis cell used for the design of the rough gratings.

\section{Characterization of roughness}

The Debye-Waller factor accounts for the damping of the scattered intensity at the orders of diffraction in the presence of roughness. The description of this factor assumes that the intensity of an undisturbed grating $I_0$ is reduced according to the amount of roughness $\xi$ and the scattering vector $q_y$. Although our reference grating is supposed to be an ideal structure that serves as a basis for the rough structures, the line shape of the different gratings may differ due to the etching process. Furthermore, the line shape of the etched structure is influenced by the distribution of the blocks and the space between the lines. Hence, some variation between the line shape of the perfect grating and that of the rough gratings is expected. Due to the high sensitivity of GISAXS, this variation influences the intensity of the diffraction orders. Therefore, the intensity of the reference grating cannot be taken as a reference $I_0$ in the equation of the Debye-Waller factor. 

Therefore, the line shape of an ideal grating must be reconstructed for each rough structure. It is only after this point that the Debye-Waller Factor can be applied. In the following, the reconstruction process is outlined together with the uncertainty evaluation. Finally, the comparison between the standard deviation of the roughness obtained with the DWF $\xi_{DWF}$ and of the design distribution $\xi_{design}$ and of the SEM images $\xi_{SEM}$ is given.

\subsection{Characterization of the line shape}

Finding the global best solution in a given parameter space is the aim of every reconstruction process. This requires an optimal measurement setup, a good describing model together with a consistent reconstruction approach, and an evaluation of the possible errors influencing those results. For the reconstruction of the line shape of the grating from the diffraction orders, an optimization based on the forward calculation of a model of the line is performed.

\begin{figure}
\centering
        \includegraphics[width=.8\linewidth]{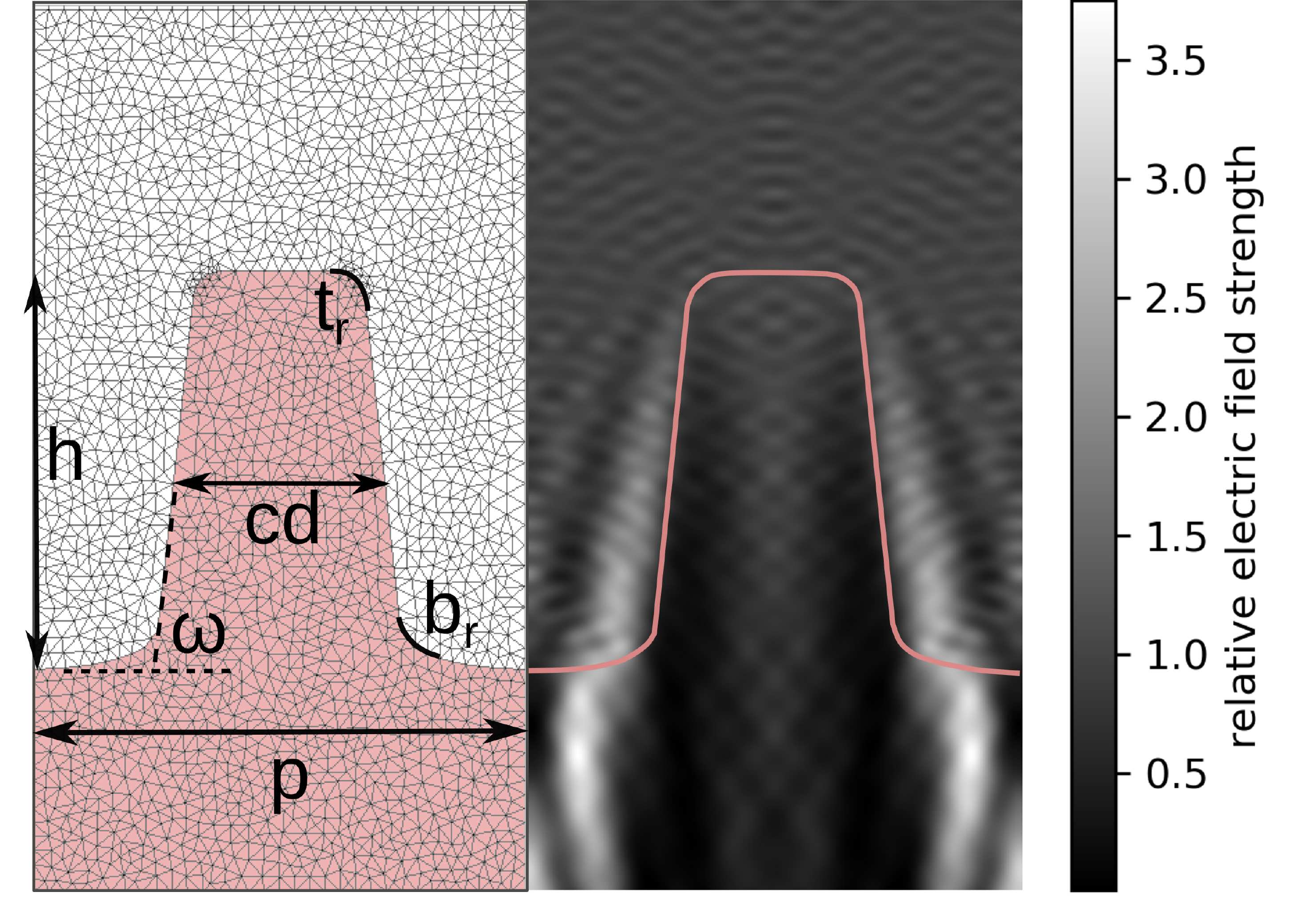}
        \caption{On the left is the computational domain with the parameterized line shape. On the right, the near-field calculation of this structure with JCMsuite is shown.}
        \label{fig:line_profile}
\end{figure}

For the computation of the diffracted intensities, different theoretical approaches can be used. One well-known approach is a Maxwell solver based on a finite element method.
An advantage to this method is that it can provide a rigorous solution of the near-field distribution of any given geometrical shape. This method has already been successfully implemented for the reconstruction of the line shape of a lamellar grating~\cite{soltwisch_reconstructing_2017}. However, it is not viable to model the line roughness with a finite element method due to the large discrepancy between the incidence wavelength of the photon beam and the size of the computational domain. Three-dimensional simulations of the line roughness would require a high discretization volume and very small discretization lengths along the propagation direction (i.e., along the lines). Therefore, instead of reconstructing a three-dimensional shape, a two-dimensional reconstruction is usually performed with the Debye-Waller factor accounting for the effect of the roughness.

We used the software package JCMsuite~\cite{pomplun_adaptive_2007} as the Maxwell solver. In Figure~\ref{fig:line_profile}, the computational domain is shown with a cross-section of the line profile and an illustrative near-field intensity distribution calculated for the line grating. For this calculation, the lines are considered infinite and periodic boundary conditions are applied to the lateral direction of the computational domain. After the computation of the  near field, the diffraction efficiencies $I_0(m)$ are obtained by means of a post-process using a Fourier transformation. Afterwards, the Debye-Waller factor is applied, yielding $I_{DWF}$. In an iterative process, these intensities are compared to the measured ones.

\begin{figure*}
     \centering
     \includegraphics[width=.8\textwidth]{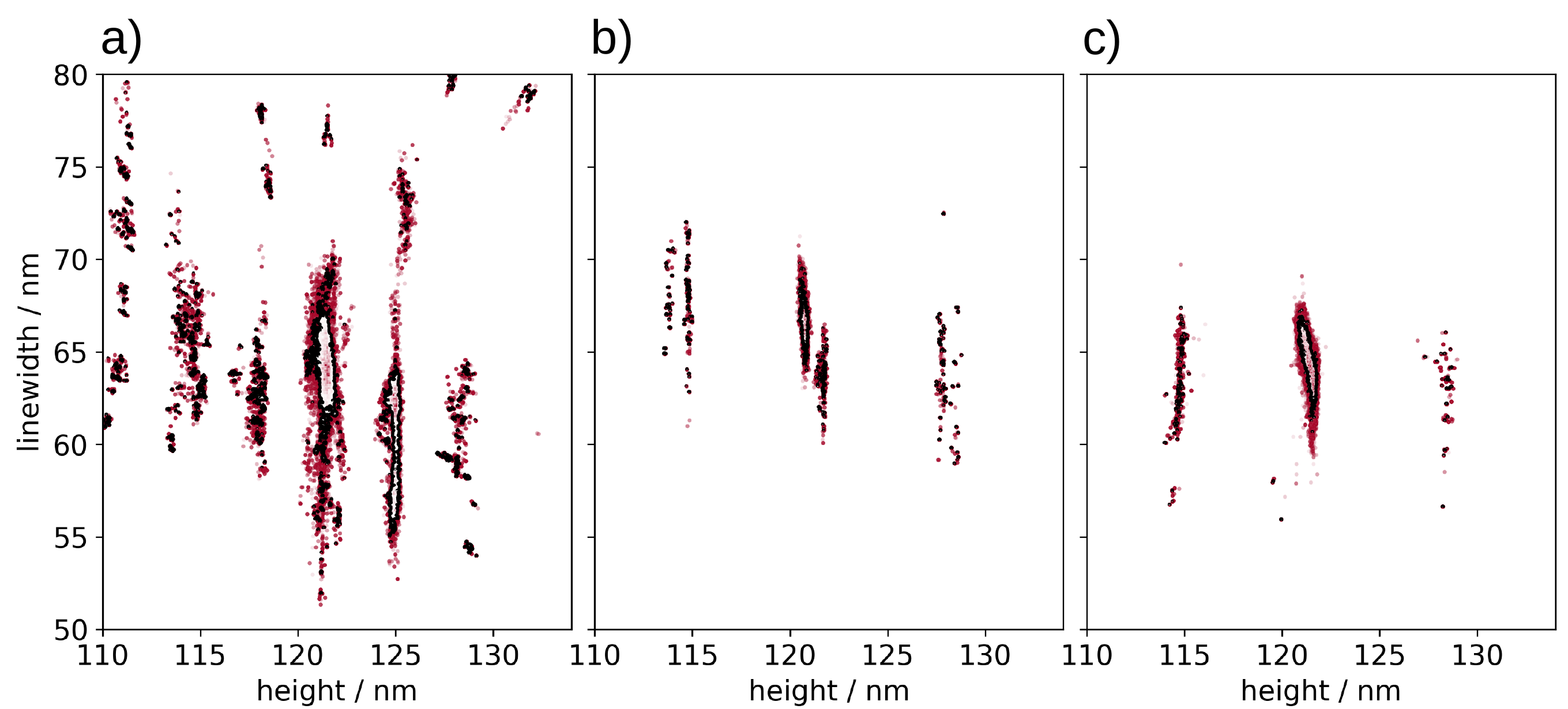}
    \caption{Comparison of the projected posterior distribution for the height and width. The black contours designate the area within 1$\sigma$. a) Distribution when only the diffraction intensities of one measurement (6keV incoming photon energy) are considered for the reconstruction. b) Three measurements with different incoming photon energy (6 keV, 6.05 keV and 6.1 keV) are considered for the optimization. c) Posterior distribution when the three measurement for different incoming photon energies and the divergence of the photon beam are considered.}
    \label{fig:ccorner}
\end{figure*}

Several optimization methods can be used~\cite{Schneider_benchmark_2018} to find a global solution of the reconstruction problem by exploring the large parameter space. However, to obtain the parameter sensitivity and the confidence intervals, a Markov Chain Monte Carlo (MCMC) sampling method~\cite{Foreman_emcee_2013} has been used. The posterior probability of the parameters depends on previous knowledge of the distribution of the parameters, which is the prior function, and on the likelihood function of the set of parameters, 

\begin{equation}
\begin{split}
posterior  \ probability  \propto\\
  likelihood \times \ prior & \ probability
\end{split}
\end{equation}

\noindent The fitted parameters are the ones that define the line-profile excluding the pitch, which is considered to be constant. The prior function is considered to be uniformly distributed for all the parameters except for the photon energy. The distribution of the photon energy is considered as a Gauss-profile with a relative width of 10$^{-4}$. The fit is performed by maximizing the likelihood (i.e., maximizing the log-likelihood), 
\begin{equation}
    \ell = \prod_{E,m} {[2\pi\sigma^2(m,E)]^{-1/2}}  \exp(-\chi^2/2),
    \label{eq:lkh}
\end{equation}
 
\noindent where $\chi^2$ corresponds to

\begin{equation}
    \chi^2= \frac{\Big(I_{DWF}\Big(m,E\Big)-I_{exp}\Big(m,E\Big)\Big)^2}{\sigma^2(m,E)},
\label{eq:chi}
\end{equation}

\noindent and $\sigma(m,E)$ comprises the uncertainty of the measurement and of the simulation. m $\in \mathbb{N}$ is the diffraction order and E the energy. A breakdown of this factor is performed in the following section.

The existence of multiple modalities in the distribution of the posterior probability is a common issue in the reconstruction process. In order to increase the measured points, several patterns were recorded using different incoming photon energies. Here, we used the incoming photon energies of 6 keV, 6.05 keV and 6.1 keV for the reconstruction. Figure~\ref{fig:ccorner} a) compares the projected posterior probability distribution when only one photon energy is considered or all three different energies b) are considered at the same time. The low dimensional projection of the posterior distribution is shown for two important geometrical parameters: the height h and the line width or critical dimension cd. The number of modalities is significantly reduced with an increased number of GISAXS scattering patterns obtained at different energies.

\begin{figure*}
    \centering
     \includegraphics[width=.8\textwidth]{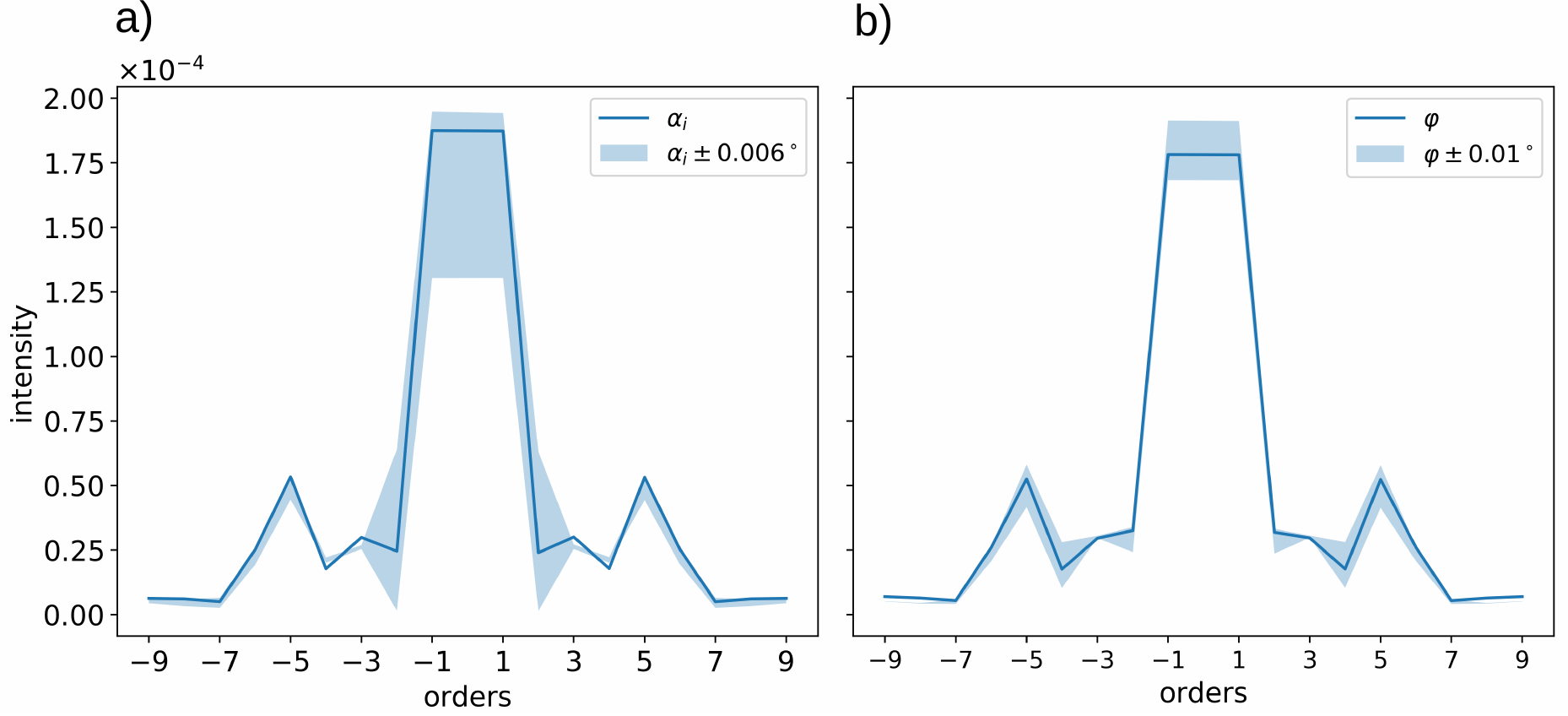}
    \caption{a) Influence of the vertical and b) horizontal divergences calculated for the line profile shown in Figure~\ref{fig:line_profile}. }
    \label{fig:div_hor_vert}
\end{figure*}

Even though the reciprocal space was mapped using several incidence photon energies, multi-modalities of the posterior distribution are still found. 
As indicated previously, the divergence of the incoming beam is $\Delta_h=0.01^\circ$ horizontally and $\Delta_v=0.006^\circ$ vertically. In spite of the fact that the values of the divergence are small, they have a considerable influence on the intensity of the measured diffraction orders (see Figure~\ref{fig:div_hor_vert}). By considering the divergence of the beam as a Gauss function weighted over the angles in the interval of the total divergence in the reconstruction process, the other possible solutions become less significant. Figure~\ref{fig:ccorner} c) shows the posterior distribution of the height and line width when the optimization is performed for three different measurements and when the divergence is considered. Comparing this approach with the other two, we can conclude that a better definition for the solution is obtained when the divergence is considered. However, even if the beam divergence is taken into account, alternative solutions are not completely ruled out.
 
\subsection{Uncertainty analysis}

The sources contributing to the uncertainties can be grouped into two different types: an experimental error $\sigma_{exp}$ and a computational error $\sigma_{comp}$. In turn, two errors known from the detector contribute to the experimental error~\cite{wernecke_characterization_2014}. Due to the Poisson statistical distribution followed by the photon counting detector, we have $\sigma_N(m, E)$, where N is the number of counts per order of diffraction m, and E is the energy. To account for the non-homogeneity of the detector, there is an additional contribution of $\sigma_{hom}(E) = 2\%$ of the measured intensities. Although this factor is energy-dependent, in this case, where the energies are close to each other, it can be considered to be the same for the three energies. Additionally, the uncertainty of the incident photon energy is taken into account by including it in the reconstruction process as a parameter with a suitable prior.

However, it was reported that the experimental error is not relevant in comparison with the computational error~\cite{soltwisch_reconstructing_2017}. The computational error, which accounts for the assumptions and approximations performed to speed up the computation, is the largest contribution to the total error. Two sources contribute to this error: a purely numerical error and another error accounting for the approximation performed in the calculation of the divergence. For the reconstruction of the line profile, the exploration of a large parameter space is needed. This would mean unacceptable calculation times when the very best solution of the solver is pursued. The precision of the calculation of the solver can be tuned by the finite element mesh size $d$ and the polynomial degree $p$ of the ansatz function to be solved. By tuning these parameters, a compromise must be found between two solutions: one that is reasonable in terms of time and one that is acceptable in terms of uncertainty. Including the effect of the divergence in the reconstruction process also entails an increase of the computation time for each function evaluation. Therefore, the number of points to weight the Gauss function must also be restricted.

Although each of these contributions can be evaluated separately, this evaluation is itself time consuming. Furthermore, this rather simple approach may disregard other error contributions. Therefore, an error model that can be included in the fit would be more effective in including all the possible error sources. The computational errors can be considered to be Gaussian distributed and energy dependent $\sigma_{comp}(E)$~\cite{soltwisch_reconstructing_2017}. An error model that was presented by Heidenreich et al.~\cite{Heidenreich_bayesian_2015} to account for the errors of a virtual scattering experiment.  Considering no systematic errors and with the error normally distributed, the variance is given by
\begin{equation}
    \sigma^2_{comp}(E) = [a(E)I(m,E)]^2 + b^2 ,
    \label{eq:error}
\end{equation}

\noindent where a(E) and b are independent. Thus, the final $\sigma(m,E)$ is

\begin{equation}
\sigma^2(m,E) = \underbrace{[a(E)I(m,E)]^2+ b^2}_{\sigma^2_{comp}(E)}  + \underbrace{\sigma^2_{N}(m,E) +\sigma^2_{hom}(E)}_{\sigma^2_{exp}(m,E)} ,
\label{eq:sigma}
\end{equation}

\noindent where $\sigma_{exp}$ is known and $\sigma_{comp}$ is included in the optimization process. If only the experimental error were considered the total error contribution of the reconstruction process would be underestimated. The factor $a(E)$ in principle accounts for the numerical error and for the approximations that are performed in order to calculate the divergence.

For the simulation, a polynomial degree $p$= 4 and a finite element size $d$ = 6 nm were chosen. For the divergence, a five-point Gauss function was considered in each direction. Although the five-point approximation could be a good mathematical approximation a priori, the orders of diffraction may be located in sensitive positions, which would lead to unexpectedly high errors. When considering the divergence, new orders of diffraction may appear at the horizon. This would result in a re-distribution of the total scattered intensity.

\subsection{Results and discussion}

The optimization was performed using the Markov Chain Monte Carlo~\cite{Foreman_emcee_2013} sampling technique with the likelihood function given in eq.~\ref{eq:lkh} and the definition of $\sigma^2$ in eq.~\ref{eq:sigma}. The posterior distribution of the optimization is analyzed after the burn-in fraction of the Markov chains for the LER1 sample. For the other samples, the MCMC technique is also performed although no uncertainty evaluation takes place to reduce the computation time. However, the confidence intervals are expected to be in the same range. 

\begin{figure}
    \centering
    \includegraphics[width=.8\linewidth]{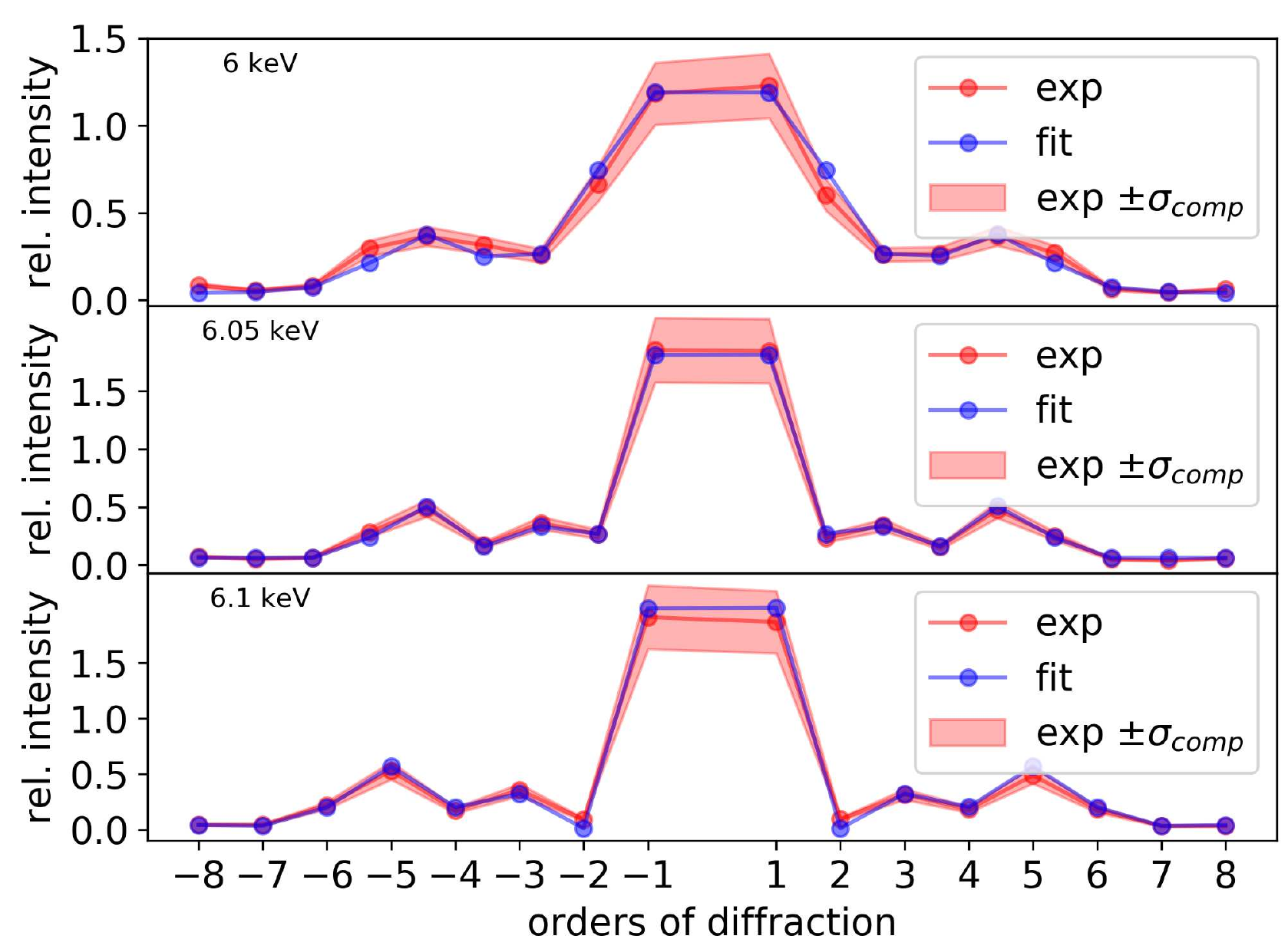}
    \caption{Comparison between the extracted intensities from the measurements (red dots) and the fit (blue dots) for the LER1 sample. The red shaded area corresponds to the fitted uncertainty.}
    \label{fig:fit_final}
\end{figure}

\subsubsection*{Multiple modalities}

In order to reduce the multiple modalities of the possible solution, additional measurement data sets and the divergence of the beam were included in the optimization. This resulted in a better defined solution although the solution was still not completely reduced to one. The probability of finding the solution around the distribution of the height at 121$\pm$ 2 nm changes with each approach, in such a way that 49$\%$ of the total samples lay under the curve when only one measurement is fitted in comparison to the fit with three measurements (63$\%$) and three energies and the divergence, where 88$\%$ of the samples are found. The issue of the multiple modalities of the posterior distribution could be solved by other means of obtaining complementary measurements of the test structure or by means of prior knowledge concerning the dimensions that are expected after the etching process~\cite{Levi_holistic_2018}. 

Considering the optimization where the divergence of the photon beam is disregarded, the fitted error contribution reaches values higher than 22 $\%$, while for the fit where the divergences are considered, the error contribution does not overtake 15$\%$. This indicates that introducing the divergence is suitable for reducing the total error of the method.

\subsubsection*{Error model}

An error model was fit (see eq.~\ref{eq:error}) resulting in values for $a(E)$ from 11.8 $\%$ for the measurement at 6 keV to 14.6$\%$ for the measreuement at 6.1 keV. There errors are much higher than the known experimental errors. 
In a parallel study, the individual contributions to the total error were studied following the procedure previously reported by Soltwisch et al.~\cite{soltwisch_reconstructing_2017}. These contributions lead to $\sigma_{comp} \approx 9\%$, which corresponds rather well to the value of 11.8$\%$ that was delivered by the MCMC. The individual analysis is close to the fitted value although some factors are still underestimated. The slightly higher MCMC-values may be due to an ignored numerical error or to an overlooked parameter in the description of the model or in the determination of the experimental error sources. Therefore, by fitting an error model, a more complete analysis of the uncertainties is delivered. By contrast, the term $b$ is negligible. It accounts for the computational background noise. As well as it could deliver the experimental background noise although this is inherently absent in a PILATUS detector.

A better error definition may also be obtained by fitting an error model that considers the dependency of the numerical error with the order of diffraction. However this would greatly increase the parameter space to be traced.

\begin{table*}
    \centering
    \caption{Fit results obtained from the scattered intensities using a Maxwell solver.}
    \label{tab:gratings}
\begin{tabular}{@{}l | c | c | c c c | c@{}}
\textrm{Parameter}& \textrm{reference}&\textrm{LER1}& \textrm{LER2}& \textrm{LWR1}& \textrm{LWR2} &  limits \\
\hline
\textrm{h/nm} & 119.50 $\pm$ 0.11 & 121.3 $\pm$ 0.3  &  122.60   &  122.93   & 123.55 & [110,134] \\  
\textrm{cd/nm} & 67.30 $\pm$ 0.31 & 64.9  $\pm$ 1.2 &  63.65   & 64.56 & 63.09  & [50,80] \\ 
\textrm{$\omega / ^\circ$} & 84.73 $\pm$ 0.33 & 83.9 $\pm$ 0.6  & 80.67  & 85.62 & 82.58  & [75,90] \\
$\xi_{design}$/ nm & - & 4.1 & 8.1 & 2.0 & 4.1 &  - \\
$\xi_{SEM}$/nm & 2.2 & 6.4 & 11.8 & 3.6 & 6.5 & - \\
$\xi_{DWF}$/nm  & 1.87  $\pm$ 0.14 & {5.11 $\pm$ 0.12} & {7.66} & {3.12 } & {5.26}  & [0,15] \\
\hline
\end{tabular}
\end{table*}

\subsubsection*{Characterization of the line}

For the determination of confidence intervals, posterior distributions were analyzed. The confidence interval of each parameter corresponds to 1$\sigma$ from the mean of the density distribution of the fits. In Table~\ref{tab:gratings}, the values obtained for the main parameters (height, cd, sidewall angle and roughness) are listed. 
The results for the reference grating have been published elsewhere ~\cite{soltwisch_reconstructing_2017}, where the vertical divergence was disregarded. The small but noticeable variation in the height of the structures across the wafer is explained by the signature of the plasma etching process. The variation in the height, together with the introduction of the line roughness, also leads to slight variation in the line profile. 
For the LER1 sample, the best fit is shown in Figure ~\ref{fig:fit_final}. There is a good agreement between the results of the optimization and the experimental data.

\subsubsection*{Roughness analysis}

\begin{figure*}
    \centering
     \includegraphics[width=\textwidth]{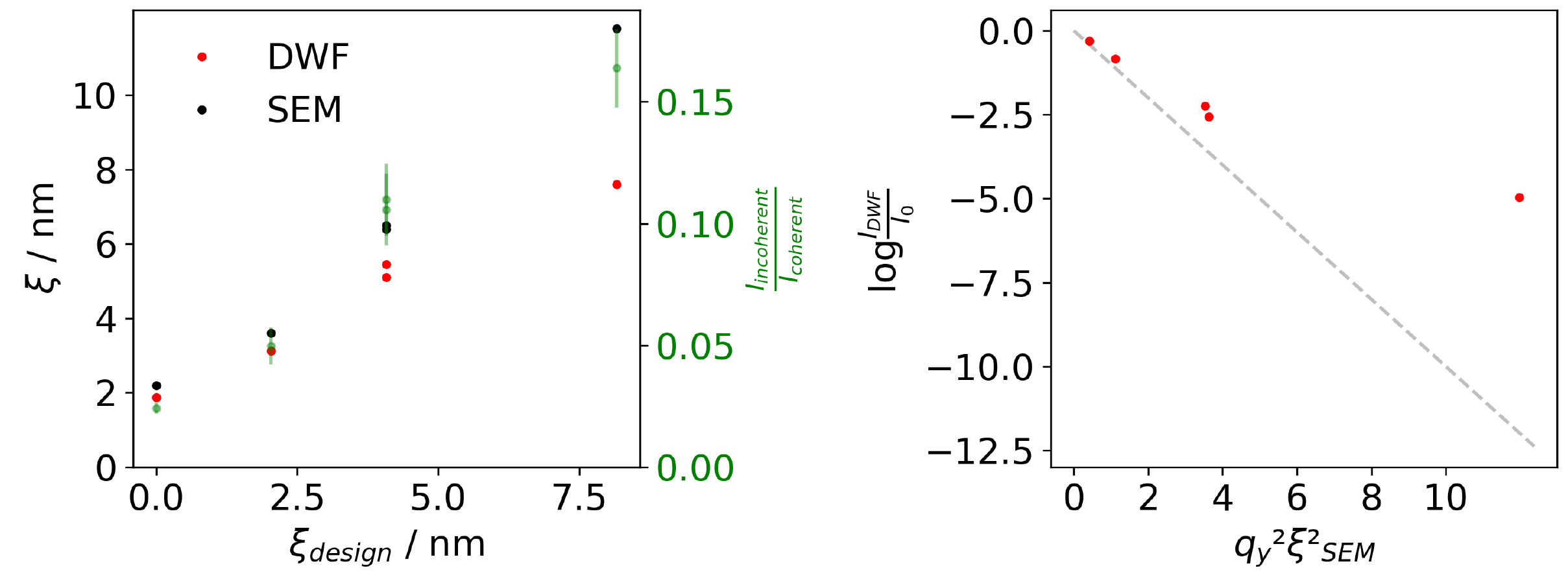}
    \caption{a) Comparison of the roughness values obtained with SEM (black) $\xi_{SEM}$ and scattering (red) $\xi_ {DWF}$ as function of the standard deviation of the roughness of the design $\xi_{design}$. In the second y-axis, the ratio between the incoherent scattered intensity and the coherent scattered intensity is shown (green). Its trend correlated with the roughness values obtained with SEM. b) Applicability of the DWF. The gray dotted line shows the expected trend of the Debye-Waller factor for the roughness values obtained by SEM. The red dots represent the intensity loss obtained for each $\xi_{SEM}$ at the highest diffraction order used (n = 7).}
    \label{fig:applicabilityDWF}
\end{figure*}

By analyzing the SEM images, the roughness distribution after the etching process was obtained. Although the design values were not met after etching, the trend of the roughness parameter was conserved. The line position follows a normal distribution instead of a uniform one. In table~\ref{tab:gratings} the standard deviation of the design distribution $\xi_{design}$ is displayed together with the standard deviation obtained using the SEM analysis $\xi_ {SEM}$ and the standard deviation obtained from the DWF $\xi_{DWF}$. For the determination of the roughness from the scattering pattern, the Debye-Waller factor was introduced to the optimization process. Figure~\ref{fig:applicabilityDWF} a) shows a comparison between the values obtained via SEM and via scattering.  The fit of the DWF (red dots) delivers smaller values than the SEM and fails to detect large amplitudes of the roughness. It should be noted, that SEM images deliver a very local value of the roughness, while in GISAXS, the whole sample is analyzed within one measurement. SEM images also have a drawback in that they provide a top picture of the structure, while in GISAXS, the average roughness over the height of the structure is given. 

In an extra y-axis, the ratio between the incoherent scattered intensity (intensity strays away from the diffraction orders) and the coherent intensity (total intensity of the diffraction orders) is shown. The measured fraction of the total incoherent scattering is limited by the area of the detector, but the contributions decay when increasing the exit angles. The data shown (green dots) corresponds to the average ratio found over twenty-one measurements that were done with different photon energies and for each grating. This ratio shows the same trend as the roughness values obtained by means of SEM. The signal scattered into the diffraction orders depends on the amplitude of the roughness, as the DWF describes. For the case of very large roughness amplitudes (that is, for the LER2) the DWF is no longer applicable (see Fig.~\ref{fig:applicabilityDWF}). The value of the roughness is underestimated by the Debye-Waller Factor. A brief derivation of the DWF is done in the appendix to discuss better the extent to which this factor faithfully describes line-edge roughness. It is shown that the DWF is applicable for Gauss distributions of the roughness or for non-Gaussian distributions when the factor $q_y^2 \xi^2$ is small enough (usually $q_y^2 \xi^2 << 1 $). However, in this specific case the DWF delivers roughness values close to the SEM values for $q_y^2\xi^2$ larger than 1 (LER1, LWR1 and LWR2 gratings), while it is not working for the distribution with the largest amplitude of the roughness (LER2 grating). There are two possible explanations for this behaviour. In one hand, the roughness distribution is close to a Gauss distribution and therefore, its impact on the scattered intensities can be evaluated with the DWF. On the other hand, several measurements were done and there is a statistically significant number of diffraction orders that are in the region or close to the region where the approximation is still applicable, independently on the distribution. With increasing values of $q_y^2\xi^2$ the discrepancies of the impact of a Gauss-distribution are larger and therefore is not longer applicable for the LER2 grating.

Additionally, it should be mentioned that for the derivation of the DWF a binary grating is considered. Hence, the effect of large variations on the height, as could result from the etching process of such a rough sample, are disregarded. We have also considered two types of roughness: LER and LWR. However, this distinction cannot be obtained from the intensity distribution of the diffraction orders. Only the observation of the diffuse scattering pattern (or more precisely, the resonant diffuse scattering sheets) can provide information on the prevailing type of roughness~\cite{Fernandez_characteristic_2017}.

\section{Conclusion}

We have analyzed the applicability of the Debye-Waller factor (DWF) for the characterization of the roughness of lamellar gratings. For the systematic analysis of the impact of the roughness in the scattering pattern, we used a set of samples with a predominant line edge roughness (LER) or line width roughness (LWR). Each line of the grating is designed to follow a uniform distribution. However, the analysis of the images obtained by means of SEM shows that the distribution of the roughness is Gaussian, due to the convolution of the pre-designed roughness with other roughness sources. 

The scattered intensities in grazing incidence small-angle x-ray scattering (GISAXS) are sensitive to small variations in the shape and to roughness. In the reconstruction process of the nanostructures, an error model was introduced to account for errors that are difficult to define in advance. For the description of the impact of the roughness on the scattered intensities, the Debye-Waller factor was used.
This factor is sensitive to variations in the roughness amplitude.
While we were able to show the validity of the DWF for values $q_y^2\xi^2 < 3 $, the DWF overestimates the actual intensity loss for the large roughness amplitudes. This is because for larger values of $q_y^2\xi^2$ the discrepancies between a Gauss and a non-Gauss distribution of the roughness become critical. The prevalent type of roughness (LER/ LWR) is observable only in the resonant diffuse scattering pattern.

In summary, the applicability of the Debye-Waller factor for non-Gaussian roughness distributions is restricted to small values of the product $q_y^2\xi^2$ (that is, $q_y^2\xi^2<1$ ). Therefore, the roughness values accessible with the DWF are limited by the number of orders of diffraction considered if the roughness is not perfectly Gauss-distributed. It is worth to mention, that this is of high importance when decreasing the sizes of the nanostructures. Nanostructures with smaller periods scatter to higher values of $q_y$. In those gratings, small variations of the roughness distribution from that of a Gaussian would prevent the use of the DWF. Therefore, further investigations of the effects of roughness in the scatter pattern are necessary.

\section{Appendix} \label{sec:app}

In Sec.~\ref{sec:DWF}, the Debye-Waller factor was briefly introduced as a common method used to account for the attenuation of the scattered signal due to the roughness. Here, for a better understanding on the applicability of the Debye-Waller factor, a short derivation of this factor for the case of line edge roughness is given. For the line width roughness, it can be done analogously.

\noindent For a perfect binary grating, the scattered intensity $I_0$ is given by~\cite{renaud_probing_2009}

\begin{equation}
    I_0 = |\sum^N_{j=0} f_j \exp(-i q_y r_j)| ^2  , 
\end{equation}

\noindent where $f_j$ is the form factor of the line at position ${r}_j $. For a grating of pitch $p$, the total length of the grating in the perpendicular direction of the lines is $Np$ and the position of each line is given by $r_j =jp$.

\noindent In the case of line edge roughness, the position along the line varies according to
\begin{equation}
    r_j = jp + \delta_{j},
\end{equation}
where $jp$ is the ideal position of the line $j$ (or the mean position of the line), $\delta_j$ is the displacement of the line and $r_j$ the new position of the edge. Considering $\delta_j$ to be independent of the original line position, we have that

\begin{equation}
 \sum^N_{j=0} f_j \exp(-i q_y j p) \Big<\exp \Big( -i (q_y \delta)\Big)\Big>
\end{equation}

The latter factor can be written as

\begin{equation}
\Big<\exp \Big( -i (q_y \delta)\Big)\Big>   = \Big< \cos(q_y \delta) \Big>   - i\Big< \sin(q_y \delta)\Big>   
\label{eq:euler}
\end{equation}

\noindent where the second term vanishes for even distributions.
\noindent In the following, we formulate the solution when two assumptions are done: for a Gauss distribution of the roughness and for small values of $q_y\xi$.

\noindent We have that the ensemble average is

\begin{equation}
    \Big< \cos(q_y \delta) \Big> = \frac{\int^{\infty}_{-\infty}
    \cos(q_y\delta)\rho(\delta) d\delta}{\int^{\infty}_{-\infty}\rho(\delta)d\delta},
    \label{eq:cos_dis}
\end{equation}

\noindent where $\rho(\delta)$ is the distribution of the roughness.

\begin{itemize}

\item  If the distribution of the roughness is considered to be Gaussian centered at 0 with a variance $\xi$, 

\begin{equation}
    \rho(\delta) = \frac{1}{\xi\sqrt{2\pi}}\exp\Big(\frac{-\delta^2}{2\xi^2}\Big).
\end{equation}

\noindent So that, the denominator in eq.~\ref{eq:cos_dis} is 1 and 

\begin{equation}
    \Big< \cos(q_y \delta) \Big> = \int^{\infty}_{-\infty}
    \cos(q_y\delta) \frac{1}{\xi\sqrt{2\pi}}\exp\Big(\frac{-\delta^2}{2\xi^2}\Big) d\delta.
    \label{eq:c}
\end{equation}

\noindent Using that~\cite{bronshtein_2007_handbook} 
\begin{equation}
    \int^{\infty}_{0}\exp(-a^2x^2) \cos(bx) dx = \frac{\sqrt\pi}{2a}\exp\Big({\frac{-b^2}{4a^2}}\Big) 
\end{equation}

\noindent we have that

\begin{equation}
    \Big< \cos(q_y \delta) \Big> = \frac{1}{\xi\sqrt{2\pi}} \frac{\xi\sqrt{2\pi}}{1} \exp\Big({\frac{-2q_y^2\xi^2}{4}}\Big)
    \label{eq:}
\end{equation}

Finally, the eq.~\ref{eq:euler} can be written for a Gauss-distribution of the roughness as 

\begin{equation}
\Big<\exp \Big( -i (q_y \delta)\Big)\Big>   =  \exp\Big({-\frac{q_y^2\xi^2}{2}}\Big)
\label{eq:pre_dwf}
\end{equation}

\item 
In the case of non-Gaussian but symmetric distributions, the eq.~\ref{eq:euler} can be reduced to the $<\cos x>$ factor. It can be expanded as a Maclaurin series, obtaining the same factor as in eq.~\ref{eq:pre_dwf} for small values of $q_y\xi$.
     
\begin{equation}
    \Big<\cos (q_y \delta)\Big> \approx 1 -  \frac{1}{2} <(q_y \delta)^2> \ + \ ... \ ,
\end{equation}

Considering $q_y\delta_y$ - or for a given $q_y$, the displacement- small, the rest of the factors are negligible. The second order term can be simplified to

\begin{equation}
    <(q_y \delta)^2> =q_y^2<\delta_j^2> =q_y^2 \xi^2 ,
\end{equation}
where $\xi$ is the variance of the distribution. 

It results in  
\begin{equation}
    \Big<\cos (q_y \delta)\Big> \approx 1 - \ \frac{1}{2} <(q_y \delta_j)^2> = 1- \frac{1}{2} q_y^2 \xi^2  \approx \exp(-\frac{1}{2}\xi^2 q_y^2).
\end{equation}
\end{itemize}

\noindent For the calculation of the scattered intensity, we have that

\begin{equation}
    I\simeq | \sum_j f_j\exp(-iq_y jp)\exp(-\frac{1}{2}\xi^2 q_y^2)|^2= I_0 \exp(-\xi^2 q_y^2),
\end{equation}

\noindent where $I_0$ is the intensity from an undisturbed grating.\\

\noindent Therefore, the DWF is applicable either for small values of $q_y\xi$ (that is, $q_y\xi<1$) or for roughness that is Gauss-distributed.

\section*{Funding}

This project has received funding from the Electronic Component Systems for European Leadership Joint Undertaking under grant agreement No 783247 - TAPES3.
This Joint Undertaking receives support from the European Union's Horizon 2020 research and innovation programme and Netherlands, France, Belgium, Germany, Czech Republic, Austria, Hungary, Israel.

\bibliography{DWF1}

\end{document}